\documentclass[preprint,5p,sort&compress]{elsarticle}

\usepackage{graphicx,epsfig,amssymb,times} 
\usepackage{amsmath,amsfonts}
\usepackage{bm}
\usepackage{epstopdf}
\usepackage[linktocpage,colorlinks]{hyperref}
\usepackage[caption=false]{subfig}
\usepackage[usenames]{color}  
\journal{Physics Letter B}
\usepackage{graphicx,epsfig,amssymb} 
\usepackage{amsmath,amsfonts, times}
\usepackage{bm} 

\usepackage{epstopdf}
\usepackage[linktocpage,colorlinks]{hyperref}
\usepackage[caption=false]{subfig}
\usepackage[usenames]{color}     
\usepackage{natbib}
\usepackage{soul}
\usepackage[utf8x]{inputenc}

\definecolor{coolblack}{rgb}{0.0, 0.18, 0.39}
\definecolor{darkred}{rgb}{0.5,0,0}
\definecolor{darkgreen}{rgb}{0,0.5,0}
\definecolor{darkblue}{rgb}{0,0,0.5}
\definecolor{lapislazuli}{rgb}{0.15, 0.38, 0.61}
\definecolor{venetianred}{rgb}{0.78, 0.03, 0.08}
\definecolor{bleudefrance}{rgb}{0.19, 0.55, 0.91}
\definecolor{dogwoodrose}{rgb}{0.84, 0.09, 0.41}
\hypersetup{colorlinks=true, citecolor=darkgreen, linkcolor=darkblue, 
	urlcolor = blue}

\def\be{\begin{equation}}
\def\ee{\end{equation}}

\newcommand{\bea}{\begin{eqnarray}}
\newcommand{\eea}{\end{eqnarray}}
\newcommand{\ben}{\begin{enumerate}}
	\newcommand{\een}{\end{enumerate}}
\newcommand{\bi}{\begin{itemize}}
	\newcommand{\ei}{\end{itemize}}

\newcommand{\thet}{{{\theta}}}

\def\ga{\mathrel{\raise.3ex\hbox{$>$\kern-.75em\lower1ex\hbox{$\sim$}}}}
\def\la{\mathrel{\raise.3ex\hbox{$<$\kern-.75em\lower1ex\hbox{$\sim$}}}}

\def\l{\left}
\def\r{\right}
\def\be{\begin{equation}}
\def\ee{\end{equation}}

\def\I_M{{I_{\scriptscriptstyle M\times M}}}

\def\be{\begin{equation}}
\def\ee{\end{equation}}
\def\bea{\begin{eqnarray}}
\def\eea{\end{eqnarray}}
\newcommand{\beq}{\begin{eqnarray}}
\newcommand{\eeq}{\end{eqnarray}}

\usepackage{graphicx,epsfig,amssymb} 
\usepackage{amsmath,amsfonts, times}
\usepackage{bm} 

\usepackage{epstopdf}
\usepackage[linktocpage,colorlinks]{hyperref}
\usepackage[caption=false]{subfig}
\usepackage[usenames]{color}     
\usepackage{natbib}
\usepackage{soul}
\usepackage[utf8x]{inputenc}

\definecolor{coolblack}{rgb}{0.0, 0.18, 0.39}
\definecolor{darkred}{rgb}{0.5,0,0}
\definecolor{darkgreen}{rgb}{0,0.5,0}
\definecolor{darkblue}{rgb}{0,0,0.5}
\definecolor{lapislazuli}{rgb}{0.15, 0.38, 0.61}
\definecolor{venetianred}{rgb}{0.78, 0.03, 0.08}
\definecolor{bleudefrance}{rgb}{0.19, 0.55, 0.91}
\definecolor{dogwoodrose}{rgb}{0.84, 0.09, 0.41}
\hypersetup{colorlinks=true, citecolor=darkgreen, linkcolor=darkblue, 
	urlcolor = blue}

\def\be{\begin{equation}}
\def\ee{\end{equation}}

\def\l{\left}
\def\r{\right}
\def\be{\begin{equation}}
\def\ee{\end{equation}}

\def\be{\begin{equation}}
\def\ee{\end{equation}}
\def\bea{\begin{eqnarray}}
\def\eea{\end{eqnarray}}

\begin{document}
\begin{frontmatter}
\title{\large On-axis scattering of scalar fields by charged rotating black holes}
\author[ufpa]{Luiz C. S. Leite}
\author[ufpasal]{Carolina L. Benone\corref{mycorrespondingauthor}}
\cortext[mycorrespondingauthor]{Corresponding author}
\ead{lben.carol@gmail.com}
\author[ufpa]{Lu\'is C. B. Crispino}
\address[ufpa]{Faculdade de F\'{\i}sica, Universidade 
Federal do Par\'a, 66075-110, Bel\'em, Par\'a, Brazil.}
\address[ufpasal]{Campus de Salin\'opolis, Universidade Federal do Par\'a,
	68721-000, Salin\'opolis, Par\'a, Brazil.}
\begin{abstract}
We investigate the scattering of a massless scalar field by a Kerr-Newman black hole, considering the case of on-axis incidence. We use the partial wave method to find numerical results for the scattering cross section, which we compare with classical and semiclassical analytical results, obtaining excellent agreement. We present a selection of plots for different values of the black hole parameters.
\end{abstract} 


\end{frontmatter}

\section{Introduction}\label{sec:int}
In the past few years, the detection of gravitational waves (GWs) has been reported by the LIGO--Virgo Collaborations~\cite{GWs,Abbott:2016nmj,Abbott:2017vtc,Abbott:2017gyy,Abbott:2017oio,TheLIGOScientific:2017qsa}. Most of the GWs signals cataloged, so far, have been produced during the merger of binary systems composed by black holes~(BHs). These detections give us compelling evidence of the existence of black holes with tens of solar masses and open the way to the observation of the strong regime of gravity. The existence of supermassive BHs has been recently supported by the Event Horizon Telescope observations, which captured the first image of a BH shadow~\cite{Akiyama:2019cqa}.

Scattering by objects plays a key role in physics. In general relativity, in particular, the measurement of light deflection led to the first experimental verification of a prevision by this theory \cite{Dyson:1920cwa, crispino2019hundred}. Moreover, by studying the scattering of null geodesics one can find the BH shadow, which can be used, for instance, to investigate deviations from the Kerr metric \cite{Bambi:2008jg,Johannsen:2010ru,Atamurotov:2013sca,Psaltis:2014mca,Psaltis:2015uza,Cunha:2015yba,Vincent:2016sjq,Cunha:2016bpi,Cunha:2016wzk,Wang:2017hjl,Cunha:2018acu} and to search for quantum fluctuations close to the event horizon \cite{Giddings:2016btb}.

The scattering of fields by compact objects gives rise to interesting effects such as glories \cite{Matzner.31.1869}, orbiting \cite{Anninos:1992ih} and rainbow scattering \cite{Dolan:2017rtj}. All these phenomena can be related to characteristics of the classical deflection function \cite{Ford:1959,newton1982scattering,nussenzveig2006diffraction,canto2013scattering}. Through a semiclassical analysis one can find analytical approximations, what helps to understand such effects, besides acting as consistency checks for the full numerical analysis.

Another key feature of scattering by BHs is superradiance \cite{Teukolsky:1974yv}, which allows low-frequency waves to experience an amplification, while the BH suffers a decrease in its parameters. This effect implies in a reflection coefficient greater than the unity, within the superradiance threshold, what accounts for a negative absorption cross section \cite{Caio:2013,PhysRevD.93.024028,PhysRevD.96.044043}.

Wave scattering by BHs has been the subject of many investigations. In particular, for scalar fields there are numerous works on the subject \cite{matzner1968scattering,Sanchez:1976fcl,sanchez1978elastic,Glampedakis:2001cx,Crispino:2009ki,PhysRevD.92.024012,Leite:2019uql}, what can be associated mainly to the simplicity of the scalar field equations compared to fermionic, electromagnetic and gravitational cases. One can see scalar fields as proxy models, used to highlight important features of scattering by compact objects. The detection of the Higgs boson, however, has brought extra experimental motivation for the study of scalar fields~\cite{Aad:2012tfa}. Moreover, there are models of dark matter which consider scalar fields with light masses \cite{Hui:2016ltb}.

The scattering of waves with spin 1/2 \cite{Dolan:2006vj,dolanthesis,Cotaescu:2014jca,Cotaescu:2016aty}, 1 \cite{Mashhoon:1973zz,Mashhoon:1974cq,Fabbri:1975sa,Crispino:2009xt} and 2 \cite{Vishveshwara:1970zz,Dolan:2007ut,Dolan:2008kf,PhysRevD.92.084056} by BHs has been considered in a number of works. For these cases we identify reminiscent behaviors of the scalar case, e.g., an oscillatory pattern coming from the orbiting and a glory pattern in the backward direction. However, when we consider non-zero spin, new effects arise, such as spin precession, polarization and helicity-reversal.

In the context of Einstein-Maxwell theory, asymptotically flat~(rotating and electrically charged) BHs are described by the Kerr-Newman solution. The evolution and stability of scalar perturbations in Kerr-Newman background have already been studied~\cite{Furuhashi:2004jk,Kokkotas:2010zd,Konoplya:2013rxa,Konoplya:2014sna,PhysRevD.98.025021}. Investigations concerning bound states formed of scalar fields~(scalar clouds) around Kerr-Newman BHs have also been carried out~\cite{PhysRevD.90.024051,PhysRevD.90.104024,PhysRevD.94.064030}. As far as we are aware of, a study of the time-independent scattering of scalar fields by Kerr-New\-man BHs is still lacking in the literature, and constitutes the aim of the present Letter. We investigate the scattering of a massless scalar field by a charged and rotating~(Kerr-Newman) BH, considering the case of a massless scalar wave impinging along the BH's rotation axis. In order to compute the (differential) scattering cross section, we consider a partial wave approach. Moreover, considering the properties of null geodesics, we compute the classical differential cross section and find a semiclassical approximation for the glory scattering, which we compare with our numerical results, finding excellent agreement. 

The remaining of this Letter is organized as follows: in Sec.~\ref{sec:scalarfield}, we review the treatment given to the massless scalar field in the Kerr-Newman background. In Sec.~\ref{sec:scattering}, we present an expression for the massless scalar scattering cross section of Kerr-Newman BHs, obtained via the partial-wave approach. We review some classical~(Subsec.~\ref{sec:geodesics}) and semiclassical~(Subsec.~\ref{sec:gloryscatt}) results of the scattering cross section, focusing in the case of Kerr-Newman BHs. Section~\ref{sec:numerical_methods} is devoted to the discussion about the numerical methods used to obtain the results presented in Sec.~\ref{sec:results}. We finish this Letter presenting some final remarks in Sec.~\ref{sec:remarks}. We adopt metric signature $(+,-,-,-)$, and natural units, such that $c=G=\hslash=1$.

\section{Scalar field in the Kerr-Newman spacetime}\label{sec:scalarfield}
In electrovacuum, BH solutions are given by the Kerr-New\-man metric, which is described by the following line element written in Boyer-Lindquist coordinates
\bea
&d s^2&= \l(1-\frac{2Mr-Q^2}{\rho^2}\r)d t^2-\frac{\rho^2}{\Delta}d r^2-\rho^2 d\theta^2 \nonumber\\
&+&2\l(\frac{2Mr-Q^2}{\rho^2}\r)a\sin^2\theta d t d\phi-\frac{\xi\sin^2\theta}{\rho^2}d\phi^2,\label{eq:linelement}
\eea
where $\Delta\equiv r^2-2Mr+a^2+Q^2$, $\rho^2\equiv r^2+a^2\cos^2\theta$, and $\xi\equiv (r^2+a^2)^2-\Delta a^2\sin^2\theta$. If the relation $a^2+Q^2<M^2$ is satisfied by the BH's mass $M$, electric charge $Q$, and rotation parameter $a$, this solution possesses two horizons, namely the event horizon, $r_+$, and the Cauchy horizon, $r_-$, which are given explicitly by $r_\pm =M\pm \sqrt{M^2-(a^2+Q^2)}$. Here, we shall be interested in such a BH spacetime.

We shall consider a massless and chargeless scalar field in this spacetime, described by the Klein-Gordon equation, which in its covariant form reads
\begin{equation}
(\sqrt{-g})^{-1}{\partial_\mu\l(\sqrt{-g} g^{\mu\nu}\partial_\nu\psi\r)}=0,\label{eq:kgeq}
\end{equation}
where $g$ and $g^{\mu \nu}$ are the determinant and the contravariant components of the metric, respectively. In order to find a solution for Eq. (\ref{eq:kgeq}), we use the following ansatz
\be
\psi_{\omega lm}(t,r,\theta,\phi)=\sum_{l=0}^{+\infty}\sum_{m=-l}^{+l}\frac{\mathcal{U}_{\omega lm}(r)}{\sqrt{r^2+a^2}}S_{\omega lm}(\theta)e^{im\phi-i\omega t},
\label{eq:fieldecomposition}
\ee
such that the radial function $\mathcal{U}_{\omega lm}$ obeys the differential equation
\be
\l(\frac{d^2}{d r_\star^2}+\mathcal{V}_{\omega lm}\r)\mathcal{U}_{\omega lm}(r_\star)=0,\label{eq:radialeq}
\ee
with
\bea
\mathcal{V}_{\omega lm}(r)&=& \l(\omega - m\frac{a}{r^2+a^2}\r)^2\nonumber\\
&+&\l[2Mr-2r^2-\Delta+\frac{3r^2}{r^2+a^2}\Delta\r]\frac{\Delta}{\l(r^2+a^2\r)^3} 
\nonumber\\
&-&\l(a^2\omega^2+\lambda_{lm}-2ma\omega\r)\frac{\Delta}{\l(r^2+a^2\r)^2}.
\label{eq:effpot}
\eea
The coordinate $r_\star$ is the tortoise coordinate, which can be found by integrating the equation
\be
\frac{dr_{\star}}{dr}=\l(\frac{r^2+a^2}{\Delta}\r).
\label{eq:tortoisecoord}
\ee

The angular solution is given by the scalar spheroidal harmonics, $S_{\omega lm}$, which satisfy the equation
\bea
&(\sin\theta)^{-1}\frac{d}{d\theta}\l(\sin\thet\frac{dS_{\omega lm}}{d\theta}\r)\nonumber\\
&+\l(\lambda_{\omega lm}+a^2\omega^2\cos^2\theta-\frac{m^2}{\sin^2\theta}\r)S_{\omega lm}=0,\label{eq:spheroidaleq}
\eea
and are normalized by
\be
\int d\theta\,\sin\theta\,\l|S_{\omega lm}(\theta)\r|^2=(2\pi)^{-1}.
\ee
Note that the angular eigenvalues~$\lambda_{lm}$ are necessary to solve both the radial~[Eq.~\eqref{eq:radialeq}] and angular~[Eq.~\eqref{eq:spheroidaleq}] equations.

Since we are dealing with a scattering problem, we consider an incoming wave from infinity, which is partially transmitted through the event horizon and partially scattered to infinity. The boundary conditions that are in agreement with these requirements are given by
\be
\mathcal{U}_{\omega lm}(r_\star)\sim\l\{
\begin{array}{c l}
	\mathcal{U}_I+{{\cal R}_{\omega lm}} \mathcal{U}_I^* & (r_\star\rightarrow +\infty),\\
	{\mathcal{T}_{\omega lm}} \mathcal{U}_T & (r_\star\rightarrow -\infty),
\end{array}\r.
\label{eq:inmodes}
\ee  
with
\be
\mathcal{U}_T = e^{-i \l({\omega-m\frac{a}{a^2+r_{+}^2}}\r)r_\star}\sum_{j=0}^N g_j (r-r_+)^j,\label{eq:horizon_series}
\ee
and
\be
\mathcal{U}_I = e^{-i \omega r_\star}\sum_{j=0}^N \frac{h_j}{r^j},\label{eq:infinity_series} 
\ee
where $g_j$ and $h_j$ are constants which can be fixed by imposing that Eq.~\eqref{eq:inmodes} obeys Eq.~\eqref{eq:radialeq}. In order to simplify our equations we assume $g_j=h_j=1$. The quantities $\mathcal{R}_{\omega lm}$ and $\mathcal{T}_{\omega lm}$ are related to the reflection and transmission coefficients, respectively.

\section{Scattering cross section} \label{sec:scattering}
Our main goal is to numerically compute the differential scattering cross section, which can be expressed as 
\be
\frac{d\sigma}{d\Omega}=|f(\theta,\phi)|^2,\label{eq:scattering_cross_section}
\ee
where $f(\theta,\phi)$ is the scattering amplitude, that for on-axis incidence~\footnote{Note that $m=0$ for on-axis incidences (see Ref.~\cite{PhysRevD.96.044043}).
} is given by the following partial wave series
\be
f(\theta)=\frac{2\pi}{i\omega}\sum_{l=0}^\infty  S_{\omega l 0}(0)S_{\omega l 0}(\theta)\left[(-1)^{l+1}\mathcal{R}_{\omega l 0}-1\right],\label{eq:partial_wave_series}
\ee
in which $\theta$ represents the scattering angle.
\subsection{Geodesics scattering} \label{sec:geodesics}
Let us first analyze the null geodesic scattering prior to show our numerical results for the scalar scattering cross section. Massless particles follow null geodesics in general relativity, which for the Kerr-Newman spacetime are described by~\cite{benoneaxis}
\bea
\rho^2 \dot{t} &=& \frac{\xi-\mathcal{L}_z a(2Mr-Q^2)}{\Delta},\label{eqt}\\
\rho^4 \dot{r}^2 &=& [(r^2+a^2)-a \mathcal{L}_z]^2 - \Delta[(\mathcal{L}_z-a)^2+\mathcal{C}],\label{eqr}\\
\rho^2 \dot{\phi} &=& \frac{a(r^2+a^2)-a^2\mathcal{L}_z}{\Delta} - \frac{a\sin^2{\theta}-\mathcal{L}_z}{\sin^2{\theta}},\label{eqphi}\\
\rho^4 \dot{\theta}^2 &=& [\mathcal{C}+(\mathcal{L}_z-a)^2] -\frac{(a\sin^2{\theta}-\mathcal{L}_z)^2}{\sin^2{\theta}},\label{eqthe}
\eea
where $\mathcal{L}_z=L_z/E$ is the azimuthal angular momentum per unit energy and $\mathcal{C}=C/E^2$ is the Carter's constant divided by the squared energy. 

For on-axis incidence, we have $\mathcal{L}_z = 0$, and the impact parameter is simply
\be
b=\sqrt{\mathcal{C}+a^2}.
\ee
Considering Eqs. (\ref{eqr}) and (\ref{eqthe}), we can find an orbital equation, namely
\be
\left(\frac{dr}{d\theta}\right)^2=\frac{(a^2+r^2)^2-b^2\Delta}{b^2-a^2 \sin{\theta}}.\label{eq:orbitaleq}
\ee
Defining $u\equiv1/r$, we can rewrite Eq.~\eqref{eq:orbitaleq} as follows: 
\be
\left(\frac{du}{d\theta}\right)^2=u^4\left[\frac{(a^2+\frac{1}{u^2})^2-b^2\Delta_u}{b^2-a^2 \sin{\theta}}\right],
\label{equ}
\ee
where $\Delta_u\equiv(1/u^2-2M/u +Q^2+a^2)$.

The deflection function can be written as~\cite{Collins:1973xf}
\be
\Theta(b)=2\beta(b)-\pi,
\label{dff}
\ee
in which
\be
\beta(b)=\int_0^{u_0}du\l\{u^4\left[\frac{(a^2+\frac{1}{u^2})^2-b^2\Delta_u}{b^2-a^2 \sin{\theta}}\right]\r\}^{-1/2},
\label{beta}
\ee
where $u_0=1/r_0$, and $r_0$ is the radius of closest approach for a null geodesic.

For a stream of particles, the classical differential scattering cross section is~\cite{Collins:1973xf}
\be
\frac{d\sigma}{d\Omega} = \frac{1}{\sin{\theta}}\sum_j b_j(\theta)\left|\frac{db_j}{d\theta}\right|,
\label{scl}
\ee
where the sum takes into account the fact that massless particles with $b$ close to the critical impact parameter~$b_c$ can orbit the BH many times before being scattered. Note that the scattering angle $\theta$ is related to the deflection angle $\Theta$ via $\Theta=\pm\theta-2N\pi$, with $N=0,1,2,\ldots$. 

\subsection{Glory scattering} \label{sec:gloryscatt}
Scattering at large angles is associated to the glory effect, which is characterized by a bright spot or halo in the backward direction~\cite{nussenzveig1969high}. Using a semiclassical approach, an analytical formula for the glory scattering in the BH case can be obtained, and is given by~\cite{DeWittMorette:1984pk,Zhang:1984vt,Matzner.31.1869,Anninos:1992ih}
\be
\l.\frac{d\sigma}{d\Omega}\r|_{\theta\approx\pi} = \mathcal{A}\,J_0^2\l(\omega b_g \sin{\theta}\r),\label{eq:glory_formula}
\ee
where $J_0$ is the Bessel function of the first kind of order 0, and we have defined the intensity of the glory peak as
\be
\mathcal{A}\equiv2\pi \omega b_g\left|\frac{db}{d\theta}\right|_{\theta=\pi},
\ee 
and the glory impact parameter $b_g\equiv b(\pi)$.

\section{Numerical methods}\label{sec:numerical_methods}
In this section we make a brief discussion about the numerical methods used to compute the scattering cross section for the massless scalar field. 

From Eqs.~\eqref{eq:scattering_cross_section} and~\eqref{eq:partial_wave_series}  we can see that the basic ingredients to compute the scattering cross section are the coefficients $\mathcal{R}_{\omega l 0}$, as well as the angular functions $S_{\omega l 0}(\theta)$ and their eigenvalues $\lambda_{\omega l0}$.

In order to obtain the coefficients $\mathcal{R}_{\omega l 0}$, we numerically integrate the radial equation, Eq.~\eqref{eq:radialeq}, using the numerical procedure described in Refs.~\cite{Caio:2013,PhysRevD.96.044043}. We point out that, for the results presented here, the series in Eqs.~\eqref{eq:horizon_series} and~\eqref{eq:infinity_series} were truncated at $N=3$. The scalar spheroidal harmonics $S_{\omega l 0}(\theta)$ and their eigenvalues $\lambda_{\omega l0}$ are obtained via spectral decomposition as described in Refs.~\cite{cook,Leite:2017zyb,PhysRevD.98.024046}. Due to a divergence at $\theta=0^{\circ}$, the series in Eq.~\eqref{eq:partial_wave_series} is poorly convergent. We can circumvent this problem using a series reduction technique, as described in~\ref{appendix:series_reduction}.
\section{Numerical results}\label{sec:results}

\begin{figure}[ht]
	\includegraphics[width=\columnwidth]{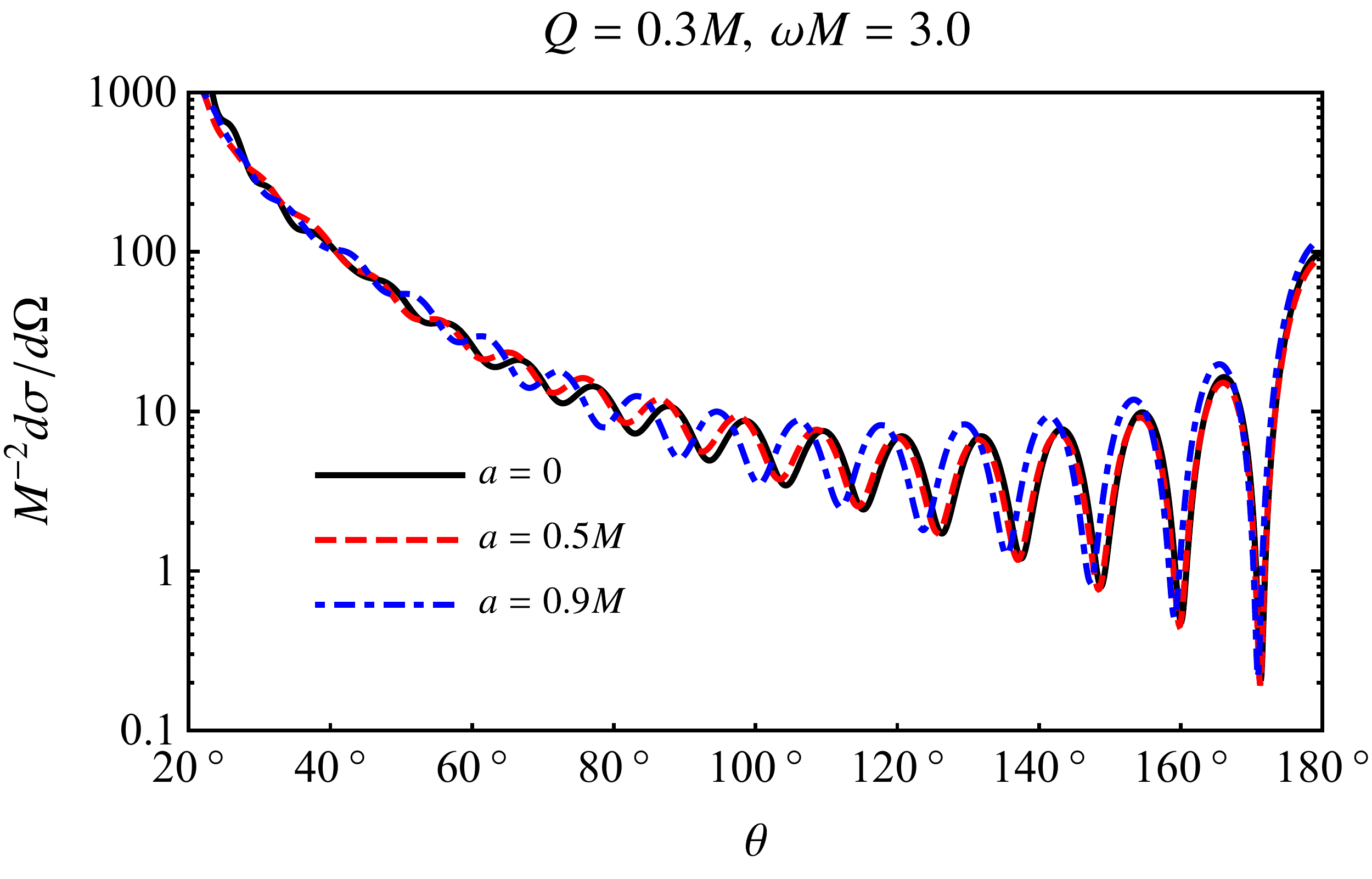}\\
	\includegraphics[width=\columnwidth]{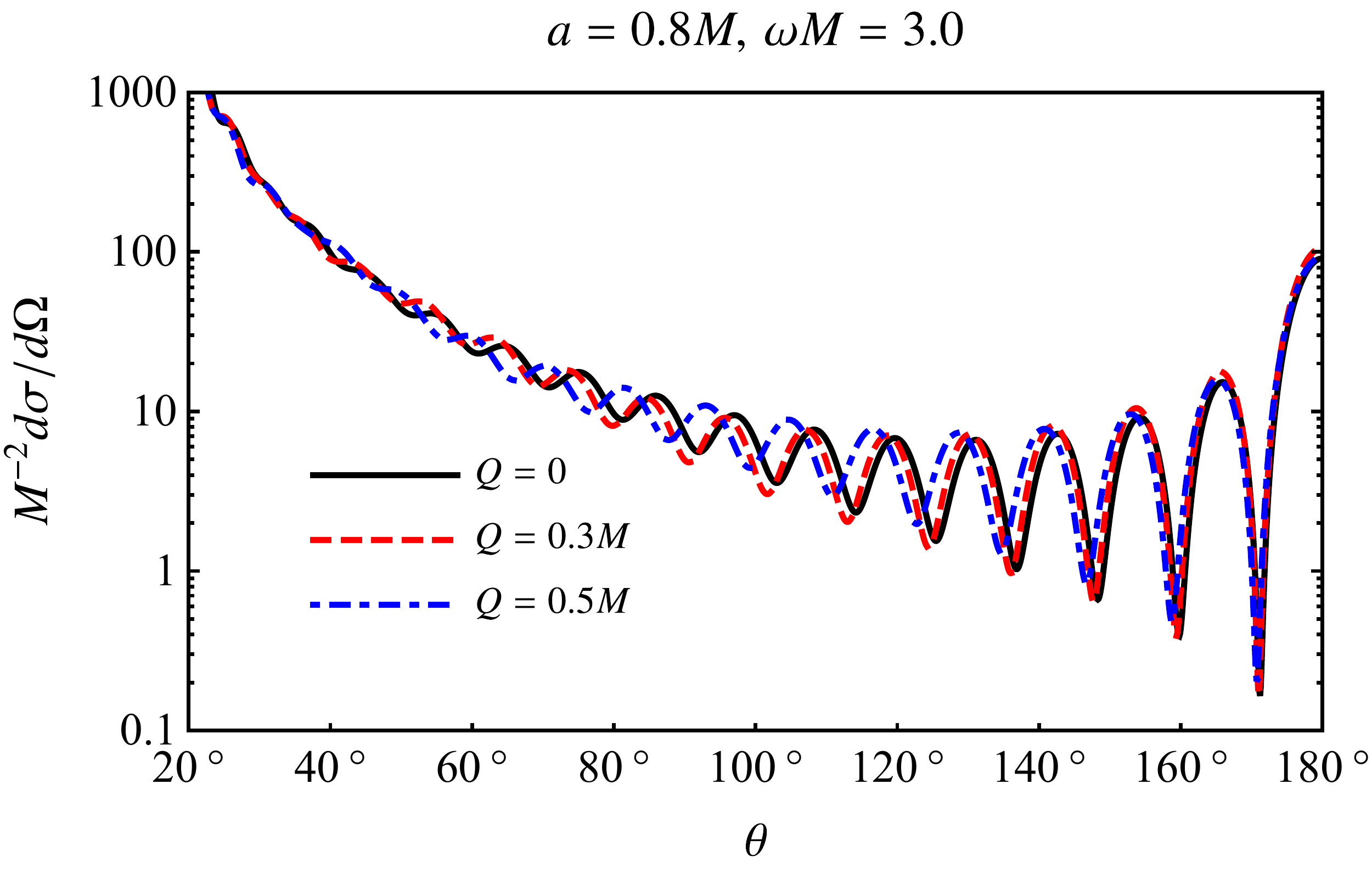}
	\caption{{\it{Top panel}}: Kerr-Newman scattering cross section for~$Q=0.3M$, and $a=0$, $0.5M$, and $0.9M$. {\it{Bottom panel}}: Massless scalar scattering cross section of Kerr-Newman BHs with rotation parameter~$a=0.8M$, and electric charges~$Q=0$, $0.3M$, and $0.5M$. In both panels we choose $\omega M=3.0$.}
	\label{fig:scatteringcs_kerrnewman}
\end{figure}

In Fig.~\ref{fig:scatteringcs_kerrnewman}, we show the massless scalar scattering cross section of Kerr-Newman BHs considering a fixed value for the electric charge~$Q=0.3M$~(top panel) and for the rotation parameter~$a=0.8M$~(bottom panel) as well. We note that the width of the oscillations become larger as we increase the value of the rotation parameter~(top panel). 
A similar behavior is observed when we consider different values of the electric charge~(bottom panel), so that as one increases the value of $Q$, the interference fringes become wider. We observe that the scattering cross section of Kerr-Newman BHs presents a high peak at $180^\circ$ (see Fig.~\ref{fig:scatteringcs_kerrnewman}), the so-called glory peak.

\begin{figure}[ht]
	\includegraphics[width=\columnwidth]{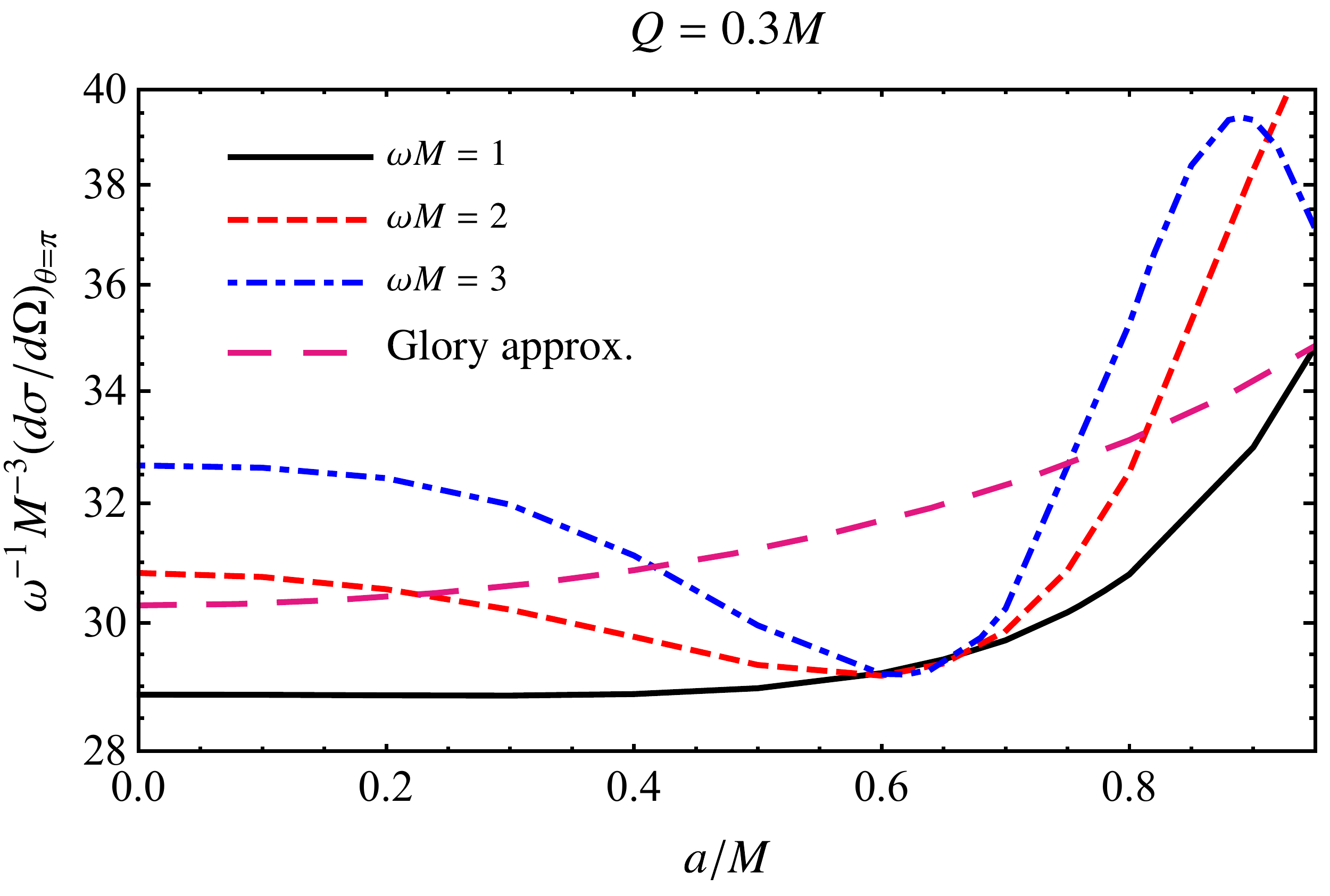}\\
	\includegraphics[width=\columnwidth]{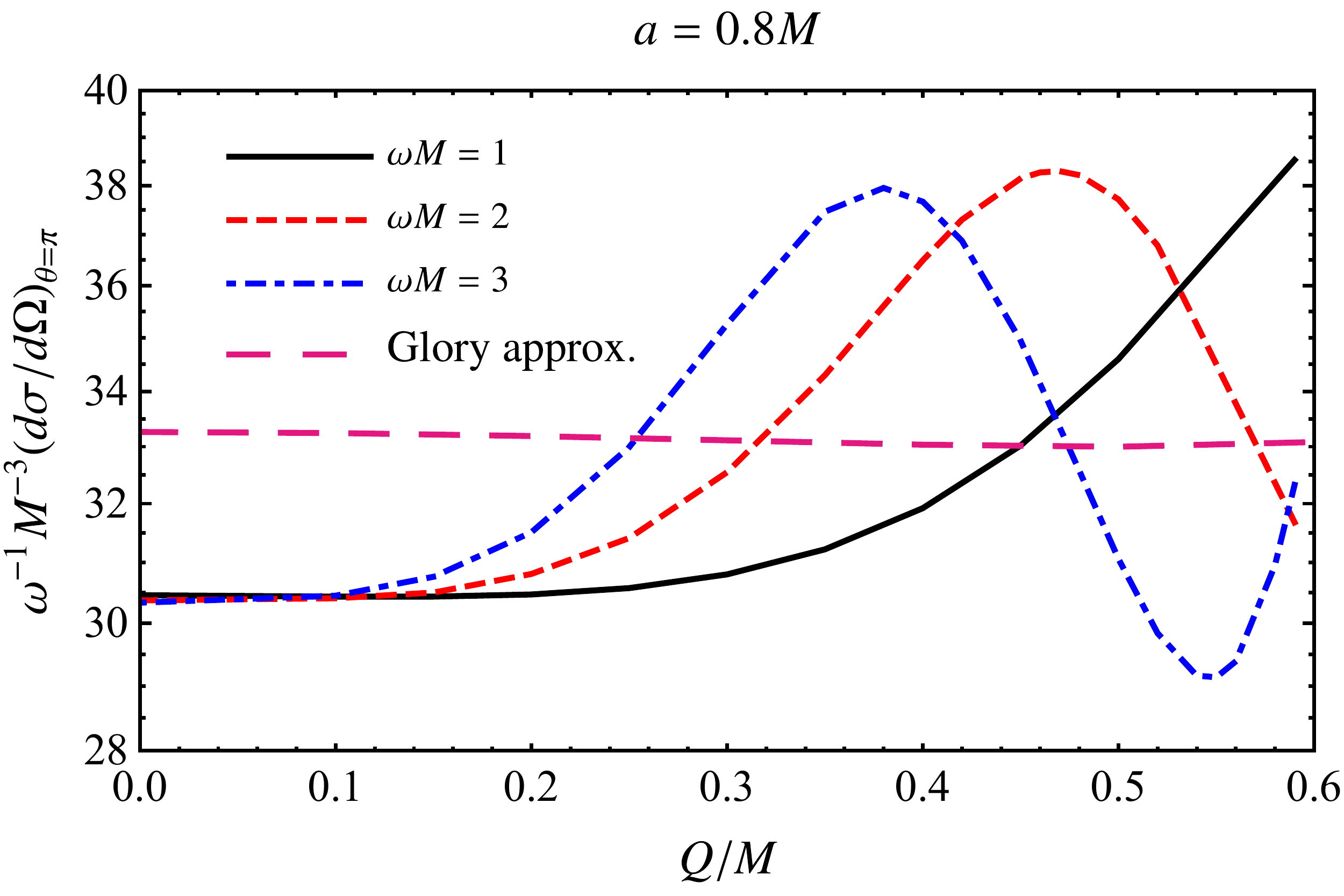}
	\caption{{\it{Top panel}}: Analytical and numerical results for the magnitude of the glory peak for Kerr-Newman BHs with $Q=0.3M$, as a function of the BH's rotation parameter. {\it{Bottom panel}}: Magnitude of the glory peak of Kerr-Newman BHs with $a=0.8M$, as a function of the BH's electric charge.}
	\label{fig:glory_peak}
\end{figure}

In Fig.~\ref{fig:glory_peak}, we exhibit the intensity of the glory peak of Kerr-Newman BHs with $Q=0.3M$~(top panel) and $a=0.8M$~(bottom panel) as a function of both the BH's rotation parameter~(top panel) and electric charge~(bottom panel), respectively. We choose different values for the wave-frequency~$\omega M=1.0$, $2.0$ and $3.0$, and we also compare our numerical results with the analytical ones obtained via the semiclassical glory scattering formula presented in Sec.~\ref{sec:gloryscatt}. We see that the glory peak exhibits a non-monotonic behavior around the semiclassical result, as we vary both the rotation parameter~(top panel) and the electric charge~(bottom panel). 

\begin{figure}[ht]
	\includegraphics[width=\columnwidth]{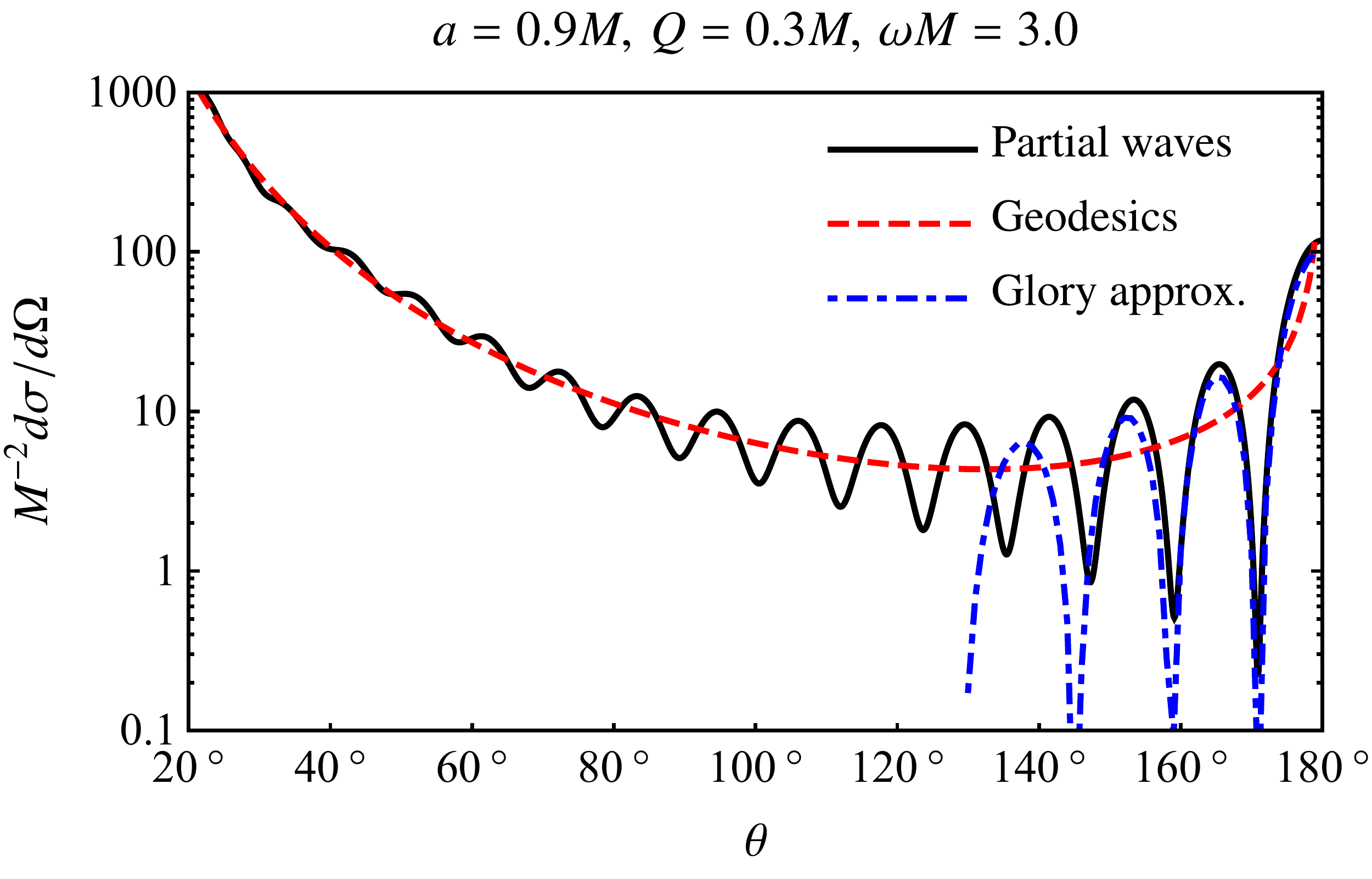}
	\caption{Comparison between the scalar scattering cross section and the classical one, for Kerr-Newman BHs with $a/M=0.9$, $Q/M=0.3$ and $M\omega =3$. We also exhibit the glory approximation.}
	\label{fig:fig_com}
\end{figure}

We finish this section by comparing our numerical results for the differential scattering cross section with the analytical ones. In Fig.~\ref{fig:fig_com}, we compare the geodesic with the scalar scattering cross sections of Kerr-Newman BHs with $a/M=0.9$, $Q/M=0.3$ and $M\omega =3.0$, together with the semiclassical glory approximation. We note that both the geodesic and the scalar scattering cross sections present a divergent behavior for small scattering angles. Moreover, the scalar scattering cross section presents a great agreement with the glory approximation near the backward direction~($\thet\approx180^{\circ}$).  

\section{Final remarks}\label{sec:remarks}
We have numerically computed the scalar scattering cross section of Kerr-Newman BHs, focusing in the case of on-axis incidence. Our numerical results have been obtained using a series reduction technique, exhibiting excellent agreement with the analytical (classical and semiclassical) ones. As in the case of Kerr BHs~\cite{Glampedakis:2001cx}, the (on-axis) cross section of a Kerr-Newman BH presents a behavior similar to the corresponding static (Reissner-Nordstrom~\cite{Crispino:2009ki}) case. 
We have shown that the angular widths of the spiral oscillations become larger as one increases the BH's rotation and/or the electric charge. We can interpret these effects through a semiclassical reasoning, as follows: from the glory formula given by Eq.~\eqref{eq:glory_formula}, the width of the glory oscillations is proportional to $1/{\omega b_g}$, and one can show that the glory impact parameter diminishes as the values of $a$ and/or $Q$ increase. We have seen that the flux scattered in the backward direction is a non-monotonic function of both the rotation parameter and the electric charge of Kerr-Newman BHs. This behavior is not revealed by the semiclassical analysis.

\section*{Acknowledgments}
The authors would like to acknowledge 
Conselho Nacional de Desenvolvimento Cient\'ifico e Tecnol\'ogico (CNPq)
and Coor\-de\-na\c{c}\~ao de Aperfei\c{c}oamento de Pessoal de N\'ivel Superior (CAPES) -- Finance Code 001, from Brazil, for partial financial support. This research has also received funding from the European Union's Horizon 2020 research and innovation programme under the H2020-MSCA-RISE-2017 Grant No. FunFiCO-777740.

\appendix
\section{Series reduction technique}\label{appendix:series_reduction}
In this Appendix, we outline the series reduction technique used to compute the massless scalar scattering cross section. This method is based mainly on Ref.~\cite{Yennie_1954:prd95_500} and has been successfully adapted to BH cases~(for an example, see Ref.~\cite{Dolan:2008kf}).

We start rewriting the partial wave series $f(\theta)$, using the spectral decomposition of $S_{\omega l 0}(\theta)$, as a sum over spherical harmonics, namely
\be
f(\theta)=\sum_{j=0}^{+\infty}F_{\omega j}Y_{j 0}(\theta),\label{eq:partialwave_series}
\ee 
where $Y_{j 0}$ are the standard scalar spherical harmonics. In Eq. \eqref{eq:partialwave_series} we have defined
\be
F_{\omega j}\equiv\frac{2\pi}{i\omega}\sum_{l=0}^{+\infty}b_{\omega jl}\, S_{\omega l 0}(0)\left[(-1)^{l+1}\mathcal{R}_{\omega l 0}-1\right],\label{eq:_series}
\ee
with $b_{\omega jl}$ being the spectral decomposition coefficients (for more details, see Refs.~\cite{cook,PhysRevD.98.024046}).

In order to improve the convergence properties of the partial wave series, one can define a new series as follows:
\be
f(\theta)=\frac{1}{(1-\cos\theta)^{n}}\sum_{j=0}^{+\infty}F{^{(n)}_{\omega j}}Y_{j 0}(\theta),\label{eq:reduced_series}
\ee 
in which $F{^{(0)}_{\omega j}}=F_{\omega j}$, with $F_{\omega j}$ given by Eq.~\eqref{eq:_series}. The reduced series $f^{(n)}(\theta)$ is less divergent at the forward direction~($\theta=0^{\circ}$). Using the properties of the spherical harmonics, one can show that the coefficients $F{^{(n+1)}_{\omega j}}$ can be obtained from the coefficients $F{^{(n)}_{\omega j}}$ via the following recurrence formula:
	\bea
	F{^{(n+1)}_{\omega j}}&=&F{^{(n)}_{\omega j}}-\l[\frac{j}{\sqrt{(2j+1)(2j-1)}}F{^{(n)}_{\omega (j-1)}}\right. \nonumber\\
	&+&\left.\frac{j+1}{\sqrt{(2j+1)(2j+3)}}F{^{(n)}_{\omega (j+1)}}\r].
	\eea

For the results presented in Sec.~\ref{sec:results}, we have computed the scattering cross section with the help of Eq.~\eqref{eq:reduced_series}, by applying three~($n=3$) iterations. Typically, we terminate the series at $l_{\rm{max}}\sim40$ and $j_{\rm{max}}\sim40$, for $\omega M=3.0$.

\bibliography{refs}

\begin{thebibliography}{10}
\expandafter\ifx\csname url\endcsname\relax
  \def\url#1{\texttt{#1}}\fi
\expandafter\ifx\csname urlprefix\endcsname\relax\def\urlprefix{URL }\fi
\expandafter\ifx\csname href\endcsname\relax
  \def\href#1#2{#2} \def\path#1{#1}\fi

\bibitem{GWs}
B.~P. Abbott, et~al.,
  \href{{http://link.aps.org/doi/10.1103/PhysRevLett.116.061102}}{{Observation
  of Gravitational Waves from a Binary Black Hole Merger}}, {Phys. Rev. Lett.}
  {116} ({2016}) {061102}.
\newblock \href {https://doi.org/{10.1103/PhysRevLett.116.061102}}
  {\path{doi:{10.1103/PhysRevLett.116.061102}}}.
\newline\urlprefix\url{{http://link.aps.org/doi/10.1103/PhysRevLett.116.061102}}

\bibitem{Abbott:2016nmj}
B.~P. Abbott, et~al., {GW151226: Observation of Gravitational Waves from a
  22-Solar-Mass Binary Black Hole Coalescence}, Phys. Rev. Lett. 116~(24)
  (2016) 241103.
\newblock \href {http://arxiv.org/abs/1606.04855} {\path{arXiv:1606.04855}},
  \href {https://doi.org/10.1103/PhysRevLett.116.241103}
  {\path{doi:10.1103/PhysRevLett.116.241103}}.

\bibitem{Abbott:2017vtc}
B.~P. Abbott, et~al., {GW170104: Observation of a 50-Solar-Mass Binary Black
  Hole Coalescence at Redshift 0.2}, Phys. Rev. Lett. 118~(22) (2017) 221101,
  [Erratum: Phys. Rev. Lett.121,no.12,129901(2018)].
\newblock \href {http://arxiv.org/abs/1706.01812} {\path{arXiv:1706.01812}},
  \href {https://doi.org/10.1103/PhysRevLett.118.221101,
  10.1103/PhysRevLett.121.129901} {\path{doi:10.1103/PhysRevLett.118.221101,
  10.1103/PhysRevLett.121.129901}}.

\bibitem{Abbott:2017gyy}
B.~P. Abbott, et~al., {GW170608: Observation of a 19-solar-mass Binary Black
  Hole Coalescence}, Astrophys. J. 851~(2) (2017) L35.
\newblock \href {http://arxiv.org/abs/1711.05578} {\path{arXiv:1711.05578}},
  \href {https://doi.org/10.3847/2041-8213/aa9f0c}
  {\path{doi:10.3847/2041-8213/aa9f0c}}.

\bibitem{Abbott:2017oio}
B.~P. Abbott, et~al., {GW170814: A Three-Detector Observation of Gravitational
  Waves from a Binary Black Hole Coalescence}, Phys. Rev. Lett. 119~(14) (2017)
  141101.
\newblock \href {http://arxiv.org/abs/1709.09660} {\path{arXiv:1709.09660}},
  \href {https://doi.org/10.1103/PhysRevLett.119.141101}
  {\path{doi:10.1103/PhysRevLett.119.141101}}.

\bibitem{TheLIGOScientific:2017qsa}
B.~Abbott, et~al., {GW170817: Observation of Gravitational Waves from a Binary
  Neutron Star Inspiral}, Phys. Rev. Lett. 119~(16) (2017) 161101.
\newblock \href {http://arxiv.org/abs/1710.05832} {\path{arXiv:1710.05832}},
  \href {https://doi.org/10.1103/PhysRevLett.119.161101}
  {\path{doi:10.1103/PhysRevLett.119.161101}}.

\bibitem{Akiyama:2019cqa}
K.~Akiyama, et~al., {First M87 Event Horizon Telescope Results. I. The Shadow
  of the Supermassive Black Hole}, Astrophys. J. 875~(1) (2019) L1.
\newblock \href {https://doi.org/10.3847/2041-8213/ab0ec7}
  {\path{doi:10.3847/2041-8213/ab0ec7}}.

\bibitem{Dyson:1920cwa}
F.~W. Dyson, A.~S. Eddington, C.~Davidson, {A Determination of the Deflection
  of Light by the Sun's Gravitational Field, from Observations Made at the
  Total Eclipse of May 29, 1919}, Phil. Trans. Roy. Soc. Lond. A 220 (1920)
  291--333.
\newblock \href {https://doi.org/10.1098/rsta.1920.0009}
  {\path{doi:10.1098/rsta.1920.0009}}.

\bibitem{crispino2019hundred}
L.~C. Crispino, D.~J. Kennefick, A hundred years of the first experimental test
  of general relativity, Nature Physics 15~(5) (2019) 416.

\bibitem{Bambi:2008jg}
C.~Bambi, K.~Freese, {Apparent shape of super-spinning black holes}, Phys. Rev.
  D79 (2009) 043002.
\newblock \href {http://arxiv.org/abs/0812.1328} {\path{arXiv:0812.1328}},
  \href {https://doi.org/10.1103/PhysRevD.79.043002}
  {\path{doi:10.1103/PhysRevD.79.043002}}.

\bibitem{Johannsen:2010ru}
T.~Johannsen, D.~Psaltis, {Testing the No-Hair Theorem with Observations in the
  Electromagnetic Spectrum: II. Black-Hole Images}, Astrophys. J. 718 (2010)
  446--454.
\newblock \href {http://arxiv.org/abs/1005.1931} {\path{arXiv:1005.1931}},
  \href {https://doi.org/10.1088/0004-637X/718/1/446}
  {\path{doi:10.1088/0004-637X/718/1/446}}.

\bibitem{Atamurotov:2013sca}
F.~Atamurotov, A.~Abdujabbarov, B.~Ahmedov, {Shadow of rotating non-Kerr black
  hole}, Phys. Rev. D88~(6) (2013) 064004.
\newblock \href {https://doi.org/10.1103/PhysRevD.88.064004}
  {\path{doi:10.1103/PhysRevD.88.064004}}.

\bibitem{Psaltis:2014mca}
D.~Psaltis, F.~Ozel, C.-K. Chan, D.~P. Marrone, {A General Relativistic Null
  Hypothesis Test with Event Horizon Telescope Observations of the black-hole
  shadow in Sgr A*}, Astrophys. J. 814~(2) (2015) 115.
\newblock \href {http://arxiv.org/abs/1411.1454} {\path{arXiv:1411.1454}},
  \href {https://doi.org/10.1088/0004-637X/814/2/115}
  {\path{doi:10.1088/0004-637X/814/2/115}}.

\bibitem{Psaltis:2015uza}
D.~Psaltis, N.~Wex, M.~Kramer, {A Quantitative Test of the No-Hair Theorem with
  Sgr A* using stars, pulsars, and the Event Horizon Telescope}, Astrophys. J.
  818~(2) (2016) 121.
\newblock \href {http://arxiv.org/abs/1510.00394} {\path{arXiv:1510.00394}},
  \href {https://doi.org/10.3847/0004-637X/818/2/121}
  {\path{doi:10.3847/0004-637X/818/2/121}}.

\bibitem{Cunha:2015yba}
P.~V.~P. Cunha, C.~A.~R. Herdeiro, E.~Radu, H.~F. Runarsson, {Shadows of Kerr
  black holes with scalar hair}, Phys. Rev. Lett. 115~(21) (2015) 211102.
\newblock \href {http://arxiv.org/abs/1509.00021} {\path{arXiv:1509.00021}},
  \href {https://doi.org/10.1103/PhysRevLett.115.211102}
  {\path{doi:10.1103/PhysRevLett.115.211102}}.

\bibitem{Vincent:2016sjq}
F.~H. Vincent, E.~Gourgoulhon, C.~Herdeiro, E.~Radu, {Astrophysical imaging of
  Kerr black holes with scalar hair}, Phys. Rev. D94~(8) (2016) 084045.
\newblock \href {http://arxiv.org/abs/1606.04246} {\path{arXiv:1606.04246}},
  \href {https://doi.org/10.1103/PhysRevD.94.084045}
  {\path{doi:10.1103/PhysRevD.94.084045}}.

\bibitem{Cunha:2016bpi}
P.~V.~P. Cunha, C.~A.~R. Herdeiro, E.~Radu, H.~F. Runarsson, {Shadows of Kerr
  black holes with and without scalar hair}, Int. J. Mod. Phys. D25~(09) (2016)
  1641021.
\newblock \href {http://arxiv.org/abs/1605.08293} {\path{arXiv:1605.08293}},
  \href {https://doi.org/10.1142/S0218271816410212}
  {\path{doi:10.1142/S0218271816410212}}.

\bibitem{Cunha:2016wzk}
P.~V.~P. Cunha, C.~A.~R. Herdeiro, B.~Kleihaus, J.~Kunz, E.~Radu, {Shadows of
  Einstein–dilaton–Gauss–Bonnet black holes}, Phys. Lett. B768 (2017)
  373--379.
\newblock \href {http://arxiv.org/abs/1701.00079} {\path{arXiv:1701.00079}},
  \href {https://doi.org/10.1016/j.physletb.2017.03.020}
  {\path{doi:10.1016/j.physletb.2017.03.020}}.

\bibitem{Wang:2017hjl}
M.~Wang, S.~Chen, J.~Jing, {Shadow casted by a Konoplya-Zhidenko rotating
  non-Kerr black hole}, JCAP 1710~(10) (2017) 051.
\newblock \href {http://arxiv.org/abs/1707.09451} {\path{arXiv:1707.09451}},
  \href {https://doi.org/10.1088/1475-7516/2017/10/051}
  {\path{doi:10.1088/1475-7516/2017/10/051}}.

\bibitem{Cunha:2018acu}
P.~V.~P. Cunha, C.~A.~R. Herdeiro, {Shadows and strong gravitational lensing: a
  brief review}, Gen. Rel. Grav. 50~(4) (2018) 42.
\newblock \href {http://arxiv.org/abs/1801.00860} {\path{arXiv:1801.00860}},
  \href {https://doi.org/10.1007/s10714-018-2361-9}
  {\path{doi:10.1007/s10714-018-2361-9}}.

\bibitem{Giddings:2016btb}
S.~B. Giddings, D.~Psaltis, {Event Horizon Telescope Observations as Probes for
  Quantum Structure of Astrophysical Black Holes}, Phys. Rev. D97~(8) (2018)
  084035.
\newblock \href {http://arxiv.org/abs/1606.07814} {\path{arXiv:1606.07814}},
  \href {https://doi.org/10.1103/PhysRevD.97.084035}
  {\path{doi:10.1103/PhysRevD.97.084035}}.

\bibitem{Matzner.31.1869}
R.~A. Matzner, C.~DeWitt-Morette, B.~Nelson, T.-R. Zhang,
  \href{http://link.aps.org/doi/10.1103/PhysRevD.31.1869}{{Glory scattering by
  black holes}}, Phys. Rev. D 31~(8) (1985) 1869--1878.
\newblock \href {https://doi.org/10.1103/PhysRevD.31.1869}
  {\path{doi:10.1103/PhysRevD.31.1869}}.
\newline\urlprefix\url{http://link.aps.org/doi/10.1103/PhysRevD.31.1869}

\bibitem{Anninos:1992ih}
P.~Anninos, C.~DeWitt-Morette, R.~A. Matzner, P.~Yioutas, T.~R. Zhang,
  {Orbiting cross-sections: Application to black hole scattering}, Phys. Rev.
  D46 (1992) 4477--4494.
\newblock \href {https://doi.org/10.1103/PhysRevD.46.4477}
  {\path{doi:10.1103/PhysRevD.46.4477}}.

\bibitem{Dolan:2017rtj}
S.~R. Dolan, T.~Stratton, {Rainbow scattering in the gravitational field of a
  compact object}, Phys. Rev. D95~(12) (2017) 124055.
\newblock \href {http://arxiv.org/abs/1702.06127} {\path{arXiv:1702.06127}},
  \href {https://doi.org/10.1103/PhysRevD.95.124055}
  {\path{doi:10.1103/PhysRevD.95.124055}}.

\bibitem{Ford:1959}
K.~W. Ford, J.~A. Wheeler, {Semiclassical description of scattering}, Annals of
  Physics 7 (1959) 259.
\newblock \href {https://doi.org/10.1016/0003-4916(59)90026-0}
  {\path{doi:10.1016/0003-4916(59)90026-0}}.

\bibitem{newton1982scattering}
R.~G. Newton, Scattering theory of waves and particles, Springer Science \&
  Business Media, 1982.

\bibitem{nussenzveig2006diffraction}
H.~M. Nussenzveig, Diffraction effects in semiclassical scattering, Vol.~1,
  Cambridge University Press, 2006.

\bibitem{canto2013scattering}
L.~F. Canto, M.~S. Hussein, {Scattering Theory of Molecules, Atoms, and
  Nuclei}, World Scientific, 2013.

\bibitem{Teukolsky:1974yv}
S.~A. Teukolsky, W.~H. Press, {Perturbations of a rotating black hole. III -
  Interaction of the hole with gravitational and electromagnetic radiation},
  Astrophys. J. 193 (1974) 443--461.
\newblock \href {https://doi.org/10.1086/153180} {\path{doi:10.1086/153180}}.

\bibitem{Caio:2013}
C.~F.~B. Macedo, L.~C.~S. Leite, E.~S. Oliveira, S.~R. Dolan, L.~C.~B.
  Crispino,
  \href{http://link.aps.org/doi/10.1103/PhysRevD.88.064033}{{Absorption of
  planar massless scalar waves by Kerr black holes}}, {Phys. Rev. D} 88 (2013)
  064033.
\newblock \href {https://doi.org/10.1103/PhysRevD.88.064033}
  {\path{doi:10.1103/PhysRevD.88.064033}}.
\newline\urlprefix\url{http://link.aps.org/doi/10.1103/PhysRevD.88.064033}

\bibitem{PhysRevD.93.024028}
C.~L. Benone, L.~C.~B. Crispino,
  \href{http://link.aps.org/doi/10.1103/PhysRevD.93.024028}{Superradiance in
  static black hole spacetimes}, Phys. Rev. D 93 (2016) 024028.
\newblock \href {https://doi.org/10.1103/PhysRevD.93.024028}
  {\path{doi:10.1103/PhysRevD.93.024028}}.
\newline\urlprefix\url{http://link.aps.org/doi/10.1103/PhysRevD.93.024028}

\bibitem{PhysRevD.96.044043}
L.~C.~S. Leite, C.~L. Benone, L.~C.~B. Crispino,
  \href{https://link.aps.org/doi/10.1103/PhysRevD.96.044043}{Scalar absorption
  by charged rotating black holes}, Phys. Rev. D 96 (2017) 044043.
\newblock \href {https://doi.org/10.1103/PhysRevD.96.044043}
  {\path{doi:10.1103/PhysRevD.96.044043}}.
\newline\urlprefix\url{https://link.aps.org/doi/10.1103/PhysRevD.96.044043}

\bibitem{matzner1968scattering}
R.~A. Matzner, {Scattering of Massless Scalar Waves by a Schwarzschild
  ``Singularity''}, Journal of Mathematical Physics 9~(1) (1968) 163--170.
\newblock \href {https://doi.org/10.1063/1.1664470}
  {\path{doi:10.1063/1.1664470}}.

\bibitem{Sanchez:1976fcl}
N.~G. Sanchez, {Scattering of scalar waves from a Schwarzschild black hole}, J.
  Math. Phys. 17~(5) (1976) 688.
\newblock \href {https://doi.org/10.1063/1.522949}
  {\path{doi:10.1063/1.522949}}.

\bibitem{sanchez1978elastic}
N.~S\'anchez, \href{http://link.aps.org/doi/10.1103/PhysRevD.18.1798}{Elastic
  scattering of waves by a black hole}, Phys. Rev. D 18 (1978) 1798--1804.
\newblock \href {https://doi.org/10.1103/PhysRevD.18.1798}
  {\path{doi:10.1103/PhysRevD.18.1798}}.
\newline\urlprefix\url{http://link.aps.org/doi/10.1103/PhysRevD.18.1798}

\bibitem{Glampedakis:2001cx}
K.~Glampedakis, N.~Andersson, {Scattering of scalar waves by rotating black
  holes}, Classical Quantum Gravity 18 (2001) 1939--1966.
\newblock \href {http://arxiv.org/abs/gr-qc/0102100}
  {\path{arXiv:gr-qc/0102100}}, \href
  {https://doi.org/10.1088/0264-9381/18/10/309}
  {\path{doi:10.1088/0264-9381/18/10/309}}.

\bibitem{Crispino:2009ki}
L.~C.~B. Crispino, S.~R. Dolan, E.~S. Oliveira, {Scattering of massless scalar
  waves by Reissner-Nordstrom black holes}, Phys. Rev. D 79 (2009) 064022.
\newblock \href {http://arxiv.org/abs/0904.0999} {\path{arXiv:0904.0999}},
  \href {https://doi.org/10.1103/PhysRevD.79.064022}
  {\path{doi:10.1103/PhysRevD.79.064022}}.

\bibitem{PhysRevD.92.024012}
C.~F.~B. Macedo, E.~S. de~Oliveira, L.~C.~B. Crispino,
  \href{http://link.aps.org/doi/10.1103/PhysRevD.92.024012}{{Scattering by
  regular black holes: Planar massless scalar waves impinging upon a Bardeen
  black hole}}, Phys. Rev. D 92 (2015) 024012.
\newblock \href {https://doi.org/10.1103/PhysRevD.92.024012}
  {\path{doi:10.1103/PhysRevD.92.024012}}.
\newline\urlprefix\url{http://link.aps.org/doi/10.1103/PhysRevD.92.024012}

\bibitem{Leite:2019uql}
L.~C.~S. Leite, C.~F.~B. Macedo, L.~C.~B. Crispino,
  \href{https://link.aps.org/doi/10.1103/PhysRevD.99.064020}{{Black holes with
  surrounding matter and rainbow scattering}}, Phys. Rev. D 99 (2019) 064020.
\newblock \href {https://doi.org/10.1103/PhysRevD.99.064020}
  {\path{doi:10.1103/PhysRevD.99.064020}}.
\newline\urlprefix\url{https://link.aps.org/doi/10.1103/PhysRevD.99.064020}

\bibitem{Aad:2012tfa}
G.~Aad, et~al., {Observation of a new particle in the search for the Standard
  Model Higgs boson with the ATLAS detector at the LHC}, Phys. Lett. B716
  (2012) 1--29.
\newblock \href {http://arxiv.org/abs/1207.7214} {\path{arXiv:1207.7214}},
  \href {https://doi.org/10.1016/j.physletb.2012.08.020}
  {\path{doi:10.1016/j.physletb.2012.08.020}}.

\bibitem{Hui:2016ltb}
L.~Hui, J.~P. Ostriker, S.~Tremaine, E.~Witten,
  \href{http://link.aps.org/doi/10.1103/PhysRevD.95.043541}{Ultralight scalars
  as cosmological dark matter}, Phys. Rev. D 95 (2017) 043541.
\newblock \href {https://doi.org/10.1103/PhysRevD.95.043541}
  {\path{doi:10.1103/PhysRevD.95.043541}}.
\newline\urlprefix\url{http://link.aps.org/doi/10.1103/PhysRevD.95.043541}

\bibitem{Dolan:2006vj}
S.~Dolan, C.~Doran, A.~Lasenby, {Fermion scattering by a Schwarzschild black
  hole}, Phys. Rev. D74 (2006) 064005.
\newblock \href {http://arxiv.org/abs/gr-qc/0605031}
  {\path{arXiv:gr-qc/0605031}}, \href
  {https://doi.org/10.1103/PhysRevD.74.064005}
  {\path{doi:10.1103/PhysRevD.74.064005}}.

\bibitem{dolanthesis}
S.~Dolan, {Scattering, absorption and emission by black holes}, Ph.D. thesis,
  Trinity Hall and Astrophysics Group, Cavendish Laboratory, University of
  Cambridge, Cambridge, England (2006).

\bibitem{Cotaescu:2014jca}
I.~I. Cotaescu, C.~Crucean, C.~A. Sporea, {Partial wave analysis of the Dirac
  fermions scattered from Schwarzschild black holes}, Eur. Phys. J. C76~(3)
  (2016) 102.
\newblock \href {http://arxiv.org/abs/1409.7201} {\path{arXiv:1409.7201}},
  \href {https://doi.org/10.1140/epjc/s10052-016-3936-9}
  {\path{doi:10.1140/epjc/s10052-016-3936-9}}.

\bibitem{Cotaescu:2016aty}
I.~I. Cotaescu, C.~Crucean, C.~Sporea, {Partial wave analysis of the Dirac
  fermions scattered from Reissner–Nordström charged black holes}, Eur.
  Phys. J. C76~(7) (2016) 413.
\newblock \href {http://arxiv.org/abs/1601.03673} {\path{arXiv:1601.03673}},
  \href {https://doi.org/10.1140/epjc/s10052-016-4260-0}
  {\path{doi:10.1140/epjc/s10052-016-4260-0}}.

\bibitem{Mashhoon:1973zz}
B.~Mashhoon, {Scattering of Electromagnetic Radiation from a Black Hole}, Phys.
  Rev. D 7 (1973) 2807--2814.
\newblock \href {https://doi.org/10.1103/PhysRevD.7.2807}
  {\path{doi:10.1103/PhysRevD.7.2807}}.

\bibitem{Mashhoon:1974cq}
B.~Mashhoon, {Electromagnetic scattering from a black hole and the glory
  effect}, Phys. Rev. D10 (1974) 1059--1063.
\newblock \href {https://doi.org/10.1103/PhysRevD.10.1059}
  {\path{doi:10.1103/PhysRevD.10.1059}}.

\bibitem{Fabbri:1975sa}
R.~Fabbri, {Scattering and absorption of electromagnetic waves by a
  Schwarzschild black hole}, Phys. Rev. D 12 (1975) 933--942.
\newblock \href {https://doi.org/10.1103/PhysRevD.12.933}
  {\path{doi:10.1103/PhysRevD.12.933}}.

\bibitem{Crispino:2009xt}
L.~C.~B. Crispino, S.~R. Dolan, E.~S. Oliveira, {Electromagnetic wave
  scattering by Schwarzschild black holes}, Phys. Rev. Lett. 102 (2009) 231103.
\newblock \href {http://arxiv.org/abs/0905.3339} {\path{arXiv:0905.3339}},
  \href {https://doi.org/10.1103/PhysRevLett.102.231103}
  {\path{doi:10.1103/PhysRevLett.102.231103}}.

\bibitem{Vishveshwara:1970zz}
C.~V. Vishveshwara, {Scattering of Gravitational Radiation by a Schwarzschild
  Black-hole}, Nature 227 (1970) 936--938.
\newblock \href {https://doi.org/10.1038/227936a0}
  {\path{doi:10.1038/227936a0}}.

\bibitem{Dolan:2007ut}
S.~R. Dolan, {Scattering of long-wavelength gravitational waves}, Phys. Rev.
  D77 (2008) 044004.
\newblock \href {http://arxiv.org/abs/0710.4252} {\path{arXiv:0710.4252}},
  \href {https://doi.org/10.1103/PhysRevD.77.044004}
  {\path{doi:10.1103/PhysRevD.77.044004}}.

\bibitem{Dolan:2008kf}
S.~R. Dolan, {Scattering and Absorption of Gravitational Plane Waves by
  Rotating Black Holes}, Classical Quantum Gravity 25 (2008) 235002.
\newblock \href {http://arxiv.org/abs/0801.3805} {\path{arXiv:0801.3805}},
  \href {https://doi.org/10.1088/0264-9381/25/23/235002}
  {\path{doi:10.1088/0264-9381/25/23/235002}}.

\bibitem{PhysRevD.92.084056}
L.~C.~B. Crispino, S.~R. Dolan, A.~Higuchi, E.~S. de~Oliveira,
  \href{http://link.aps.org/doi/10.1103/PhysRevD.92.084056}{Scattering from
  charged black holes and supergravity}, Phys. Rev. D 92 (2015) 084056.
\newblock \href {https://doi.org/10.1103/PhysRevD.92.084056}
  {\path{doi:10.1103/PhysRevD.92.084056}}.
\newline\urlprefix\url{http://link.aps.org/doi/10.1103/PhysRevD.92.084056}

\bibitem{Furuhashi:2004jk}
H.~Furuhashi, Y.~Nambu, {Instability of massive scalar fields in Kerr-Newman
  space-time}, Prog. Theor. Phys. 112 (2004) 983--995.
\newblock \href {http://arxiv.org/abs/gr-qc/0402037}
  {\path{arXiv:gr-qc/0402037}}, \href {https://doi.org/10.1143/PTP.112.983}
  {\path{doi:10.1143/PTP.112.983}}.

\bibitem{Kokkotas:2010zd}
K.~D. Kokkotas, R.~A. Konoplya, A.~Zhidenko, {Quasinormal modes, scattering and
  Hawking radiation of Kerr-Newman black holes in a magnetic field}, Phys. Rev.
  D83 (2011) 024031.
\newblock \href {http://arxiv.org/abs/1011.1843} {\path{arXiv:1011.1843}},
  \href {https://doi.org/10.1103/PhysRevD.83.024031}
  {\path{doi:10.1103/PhysRevD.83.024031}}.

\bibitem{Konoplya:2013rxa}
R.~A. Konoplya, A.~Zhidenko, {Massive charged scalar field in the Kerr-Newman
  background I: quasinormal modes, late-time tails and stability}, Phys. Rev.
  D88 (2013) 024054.
\newblock \href {http://arxiv.org/abs/1307.1812} {\path{arXiv:1307.1812}},
  \href {https://doi.org/10.1103/PhysRevD.88.024054}
  {\path{doi:10.1103/PhysRevD.88.024054}}.

\bibitem{Konoplya:2014sna}
R.~A. Konoplya, A.~Zhidenko, {Massive charged scalar field in the Kerr-Newman
  background: Hawking radiation}, Phys. Rev. D89~(8) (2014) 084015.
\newblock \href {http://arxiv.org/abs/1402.1998} {\path{arXiv:1402.1998}},
  \href {https://doi.org/10.1103/PhysRevD.89.084015}
  {\path{doi:10.1103/PhysRevD.89.084015}}.

\bibitem{PhysRevD.98.025021}
Y.~Huang, D.-J. Liu, X.-h. Zhai, X.-z. Li,
  \href{https://link.aps.org/doi/10.1103/PhysRevD.98.025021}{{Instability for
  massive scalar fields in Kerr-Newman spacetime}}, Phys. Rev. D 98 (2018)
  025021.
\newblock \href {https://doi.org/10.1103/PhysRevD.98.025021}
  {\path{doi:10.1103/PhysRevD.98.025021}}.
\newline\urlprefix\url{https://link.aps.org/doi/10.1103/PhysRevD.98.025021}

\bibitem{PhysRevD.90.024051}
S.~Hod, \href{https://link.aps.org/doi/10.1103/PhysRevD.90.024051}{{Kerr-Newman
  black holes with stationary charged scalar clouds}}, Phys. Rev. D 90 (2014)
  024051.
\newblock \href {https://doi.org/10.1103/PhysRevD.90.024051}
  {\path{doi:10.1103/PhysRevD.90.024051}}.
\newline\urlprefix\url{https://link.aps.org/doi/10.1103/PhysRevD.90.024051}

\bibitem{PhysRevD.90.104024}
C.~L. Benone, L.~C.~B. Crispino, C.~Herdeiro, E.~Radu,
  \href{https://link.aps.org/doi/10.1103/PhysRevD.90.104024}{Kerr-newman scalar
  clouds}, Phys. Rev. D 90 (2014) 104024.
\newblock \href {https://doi.org/10.1103/PhysRevD.90.104024}
  {\path{doi:10.1103/PhysRevD.90.104024}}.
\newline\urlprefix\url{https://link.aps.org/doi/10.1103/PhysRevD.90.104024}

\bibitem{PhysRevD.94.064030}
Y.~Huang, D.-J. Liu,
  \href{https://link.aps.org/doi/10.1103/PhysRevD.94.064030}{{Scalar clouds and
  the superradiant instability regime of Kerr-Newman black hole}}, Phys. Rev. D
  94 (2016) 064030.
\newblock \href {https://doi.org/10.1103/PhysRevD.94.064030}
  {\path{doi:10.1103/PhysRevD.94.064030}}.
\newline\urlprefix\url{https://link.aps.org/doi/10.1103/PhysRevD.94.064030}

\bibitem{benoneaxis}
C.~L. Benone, L.~C.~S. Leite, L.~C.~B. Crispino, S.~R. Dolan, {On-axis scalar
  absorption cross section of Kerr--Newman black holes: Geodesic analysis, sinc
  and low-frequency approximations}, International Journal of Modern Physics D
  27~(11) (2018) 1843012.
\newblock \href {https://doi.org/10.1142/S0218271818430125}
  {\path{doi:10.1142/S0218271818430125}}.

\bibitem{Collins:1973xf}
P.~A. Collins, R.~Delbourgo, R.~M. Williams, {On the elastic Schwarzschild
  scattering cross-section}, J. Phys. A 6 (1973) 161--169.
\newblock \href {https://doi.org/10.1088/0305-4470/6/2/007}
  {\path{doi:10.1088/0305-4470/6/2/007}}.

\bibitem{nussenzveig1969high}
H.~M. Nussenzveig, {High-frequency scattering by a transparent sphere. II.
  Theory of the rainbow and the glory}, {Journal of Mathematical Physics}
  10~(1) (1969) 125--176.
\newblock \href {https://doi.org/10.1063/1.1664747}
  {\path{doi:10.1063/1.1664747}}.

\bibitem{DeWittMorette:1984pk}
C.~DeWitt-Morette, B.~L. Nelson, {Glories---and other degenerate points of the
  action}, Phys. Rev. D 29 (1984) 1663--1668.
\newblock \href {https://doi.org/10.1103/PhysRevD.29.1663}
  {\path{doi:10.1103/PhysRevD.29.1663}}.

\bibitem{Zhang:1984vt}
T.~R. Zhang, C.~DeWitt-Morette, {WKB Cross Section for Polarized Glories of
  Massless Waves in Curved Space-Times}, Phys. Rev. Lett. 52 (1984) 2313--2316.
\newblock \href {https://doi.org/10.1103/PhysRevLett.52.2313}
  {\path{doi:10.1103/PhysRevLett.52.2313}}.

\bibitem{cook}
G.~B. Cook, M.~Zalutskiy,
  \href{https://link.aps.org/doi/10.1103/PhysRevD.90.124021}{{Gravitational
  perturbations of the Kerr geometry: High-accuracy study}}, Phys. Rev. D 90
  (2014) 124021.
\newblock \href {https://doi.org/10.1103/PhysRevD.90.124021}
  {\path{doi:10.1103/PhysRevD.90.124021}}.
\newline\urlprefix\url{https://link.aps.org/doi/10.1103/PhysRevD.90.124021}

\bibitem{Leite:2017zyb}
L.~C.~S. Leite, S.~R. Dolan, L.~C.~B. Crispino, {Absorption of electromagnetic
  and gravitational waves by Kerr black holes}, Phys. Lett. B774 (2017)
  130--134.
\newblock \href {http://arxiv.org/abs/1707.01144} {\path{arXiv:1707.01144}},
  \href {https://doi.org/10.1016/j.physletb.2017.09.048}
  {\path{doi:10.1016/j.physletb.2017.09.048}}.

\bibitem{PhysRevD.98.024046}
L.~C.~S. Leite, S.~Dolan, L.~C.~B. Crispino,
  \href{https://link.aps.org/doi/10.1103/PhysRevD.98.024046}{Absorption of
  electromagnetic plane waves by rotating black holes}, Phys. Rev. D 98 (2018)
  024046.
\newblock \href {https://doi.org/10.1103/PhysRevD.98.024046}
  {\path{doi:10.1103/PhysRevD.98.024046}}.
\newline\urlprefix\url{https://link.aps.org/doi/10.1103/PhysRevD.98.024046}

\bibitem{Yennie_1954:prd95_500}
D.~R. Yennie, D.~G. Ravenhall, R.~N. Wilson,
  \href{http://link.aps.org/doi/10.1103/PhysRev.95.500}{{Phase-Shift
  Calculation of High-Energy Electron Scattering}}, Phys. Rev. 95~(2) (1954)
  500--512.
\newblock \href {https://doi.org/10.1103/PhysRev.95.500}
  {\path{doi:10.1103/PhysRev.95.500}}.
\newline\urlprefix\url{http://link.aps.org/doi/10.1103/PhysRev.95.500}

\end{thebibliography}
\end{document}